# Electrically driven reprogrammable vanadium dioxide metasurface using binary control for broadband beam-steering

Matthieu Proffit[1], Sara Peliviani[1], Pascal Landais[2] and A. Louise Bradley[1,3*]

[1] School of Physics and AMBER, Trinity College Dublin, Dublin 2, Ireland

[2] School of Electronic Engineering, Dublin City University, Glasnevin, Dublin 9, Ireland

[3] IPIC, Tyndall National Institute, Cork, Ireland

ABSTRACT: Resonant optical phased arrays are a promising way to reach fully reconfigurable metasurfaces in the optical and NIR regimes with low energy consumption, low footprint and high reliability. Continuously tunable resonant structures suffer from inherent drawbacks such as low phase range, amplitude-phase correlation or extreme sensitivity that makes precise control at the individual element level very challenging. In order to bypass these issues, we use 1-bit (binary) control for beam steering for an innovative nano-resonator antenna and explore the theoretical capabilities of such phased arrays. A thermally realistic structure based on vanadium dioxide sandwiched in a metal-insulator-metal structure is proposed and optimized using inverse design to enhance its performance at 1550 nm. Continuous beam steering over 90° range is successfully achieved using binary control, with excellent agreement with predictions based on the theoretical first principles description of phased arrays. Furthermore a broadband response from 1500 nm to 1700 nm is achieved. The robustness of the design manufacturing imperfections is also demonstrated. This simplified approach can be implemented to optimize tunable nanophotonic phased array metasurfaces based on other materials or phased sifting mechanisms for various functionalities.

Nano-sized phased arrays are investigated for various near-infrared or optical applications such as flat optics[1], LIDAR[2] or optical communications[3]. They require sub-wavelength control which means nano-antennas capable of phase modulation must be engineered and manufactured. The transfer of phased arrays from RF/mmW ranges to the optical domain[4] would enable a drastic reduction in cost, size, complexity, reliability and energy consumption for LIDAR systems[2] but also enable a new generation of 2D reconfigurable optical elements[5].

One of the main ways to achieve this is by using the phase shift which occurs in a resonant antenna where light and matter strongly couple[6–9]. In order to tune the resonance, several parameters can be changed, many designs modify a geometric feature to achieve phase control with their response fixed at fabrication[10,11]. Some designs are switchable and possess two operating states[12–14] but to achieve a fully reconfigurable device with a large number of degrees of freedom, each antenna must be individually controlled post-fabrication[15]. For a given geometry, one can change the material properties using the electro-optic effect[6], carrier doping[15–19], thermo-optic effect[20] or phase change materials like germanium-antimony-tellurium alloys (GST)[21–23] or vanadium dioxide ($VO_2$). GST-based resonant metasurfaces have been experimentally tested[24] but the difficulties of experimentally changing its material have only recently been partially lifted[25]. $VO_2$ is one of the most promising materials to achieve phase change due to its significant change in optical properties and relative ease to trigger the material transition. Its insulator to metal transition (IMT) occurs around 68°C over a range of temperature[26,27] wherein a mix of the two phases coexists to constitute an intermediate material. The structural phase change of vanadium dioxide from dielectric to metallic around 68°C enables the development of tunable nanostructures for amplitude, polarization or phase control. To this date, several $VO_2$-based metasurface designs have been investigated by simulations[28] and experiments[8,29]. A continuous phase shift of up to 250 degrees at approximately 1550 nm has been achieved experimentally by thermally tuning a $VO_2$ nano-antenna array[29]. Many challenges remain in achieving this performance at the individual antenna

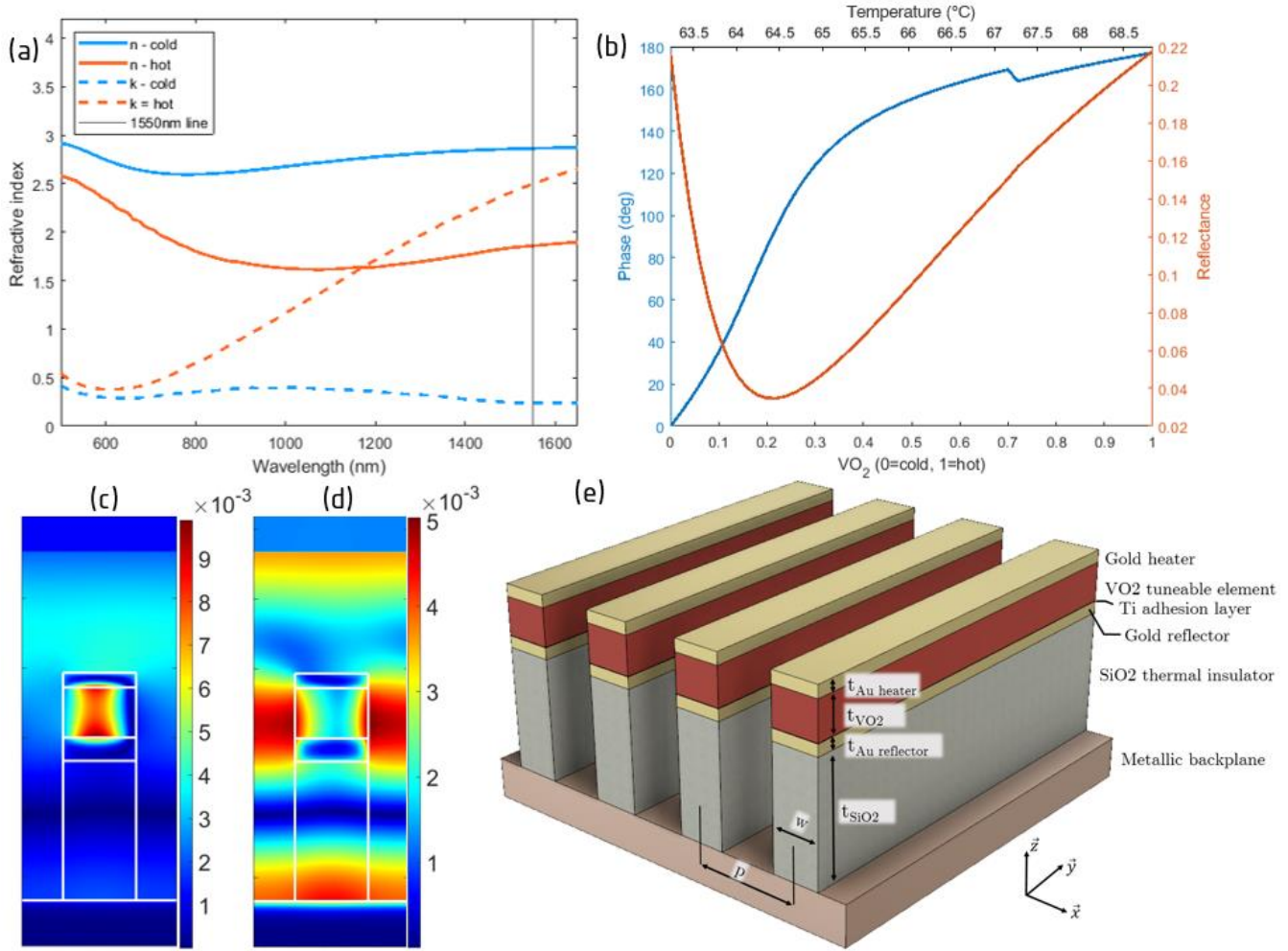

Figure 1: a - Optical properties of $VO_2$ as a function of wavelength in cold and hot states. This data was obtained using spectroscopic ellipsometry and is used for the FDTD simulations [see ref 33]   b – The dependence of the output phase and reflectance at 1550 nm on the volume fraction of $VO_2$ in the hot state (which is proportional to the material temperature, see top x-axis) for a metasurface of identical antennas and an angle of incidence of 45°   c/d – FDTD simulation at 1550 nm of the magnetic field $H_y$ field in the antenna in its dielectric cold state and metallic hot state, respectively   e – 3D structure schematic with design variables, materials and coordinate system

scale. For example, one can mention mitigating thermal crosstalk between elements which prevents individual control or extending the phase shift range up to 2π. Furthermore, the resonant nature of the device poses two problems: the amplitude variations are not easily uncorrelated from the phase shift[15,30] and the phase shift varies very abruptly and non-linearly[19] with temperature which makes precise control of each element very challenging. It is possible to mitigate these limitations[31] at the cost of other performance indicators such as the maximum achievable phase shift or reflectance but they remain intrinsically linked to the resonant nature of the antennas.

Most of these difficulties arise from the fact that a continuous phase shift such as that implemented in radio-frequency phased arrays is targeted. However, what is feasible at the macro-scale in the RF range may not be realistically applicable at the nano-scale. We propose to simplify the continuous phase shifting by using 1-bit (binary) control of the array, the complexity and most problems associated with resonant antennas are drastically reduced whilst control over the far-field amplitude pattern is retained.

In this paper, we introduce a metasurface based on a metal-insulator-metal (MIM) structure, which includes a layer of $VO_2$ as the tunable component. We use this example without loss of generality to consider a theoretical analysis of a binary controlled phased array metasurface, we demonstrate excellent properties for beam steering applications can be achieved. A continuum of anomalous refraction angles can be obtained over a wide angular range and the beam shape and width do not differ from the continuous phase shift case. Binary control can be applied to metasurfaces comprised of tunable antenna based of other materials and tuning mechanisms. We then optimize the individual MIM antenna for binary control in a $VO_2$ metasurface using inverse design, an emphasis is put on its thermal behavior both at the antenna and the array scale to ensure our design is tunable using Joule heating. This antenna is shown to have excellent robustness regarding manufacturing inaccuracies and broadband response (1500-1700nm) is achieved. We finally carry out FDTD simulations to assess the performance of this nano-antenna design in an array and successfully demonstrate beam steering with an excellent agreement with theory.

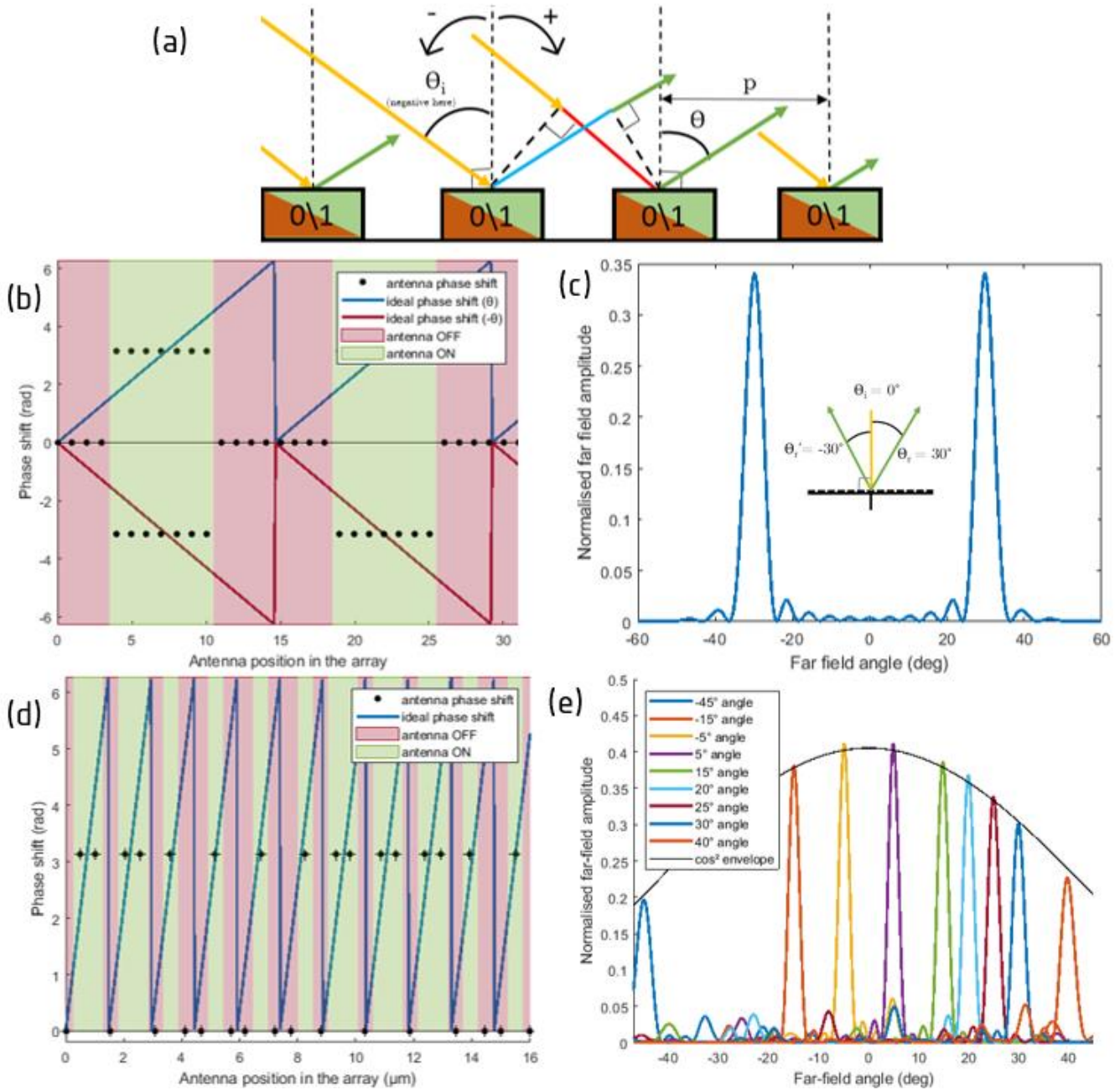

Figure 2: a – Schematic of a sample binary phased array   b - Illustration of the "-θ problem": the binary control algorithm cannot distinguish between beam steering at $\theta_r$ and $-\theta_r$ as a phase shift of $\pi$ and $-\pi$ corresponds to the same physical action   c – Far field pattern obtained with normal incident light and $\theta_r=30°$. Inset: schematic of the array configuration   d – Ideal phase profile and the corresponding antenna states with binary control for $\theta_i=45°$ and $\theta_r=20°$. This results in an aperiodic arrangement of ON and OFF antennas along the array   e – Far field pattern calculated for a fixed angle of incidence $\theta_i=45°$, N=64 antennas and using binary control to achieve various values of $\theta_r$. The cos² envelope corresponding to the antenna factor is shown to explain the lower amplitude at higher angles of anomalous reflection.

Section I: Resonant antenna using $VO_2$

Vanadium dioxide is a material that exhibits a volatile structural change over a temperature range around 68°C[32]. A transition from a monoclinic arrangement to a tetragonal rutile structure occurs which results in a drastic change in the complex refractive index as shown in Figure 1a[33], especially in the infrared spectral range. This high material property modulation enables optical tuning with very little power consumption unlike other phenomena such as the thermo-optic effect which is much smaller in magnitude and requires higher temperature modulation to achieve a meaningful change in material properties. It has long been debated which phenomenon of the structural change (Peirls distortion) or the Mott insulator behavior was responsible for the large index change of $VO_2$ as they happen almost simultaneously[34]. The $VO_2$ transition can be triggered in many ways, including ultrafast optical excitation[35], stress[36], strain[34], thermal excitation[17,30], or electric current[37]. For beam steering, the simplest approach based on Joule heating is exploited. By applying a current in the gold heater (see Figure 1d), the temperature is locally raised to switch the $VO_2$ phase. Indirectly using current to trigger $VO_2$ transition avoids filamentation[37] which occurs when a $VO_2$ element is directly subjected to a voltage. This percolative phenomenon is less reliable and presents limitations for implementation. The metasurface is comprised of a MIM antenna with a period of $\lambda/3$, a thermal insulation layer and a conductive backplane. The layer thicknesses are given in Table 2 and are the result of the thermal model and inverse design model which are detailed later in the paper. The phase and reflectance of the $VO_2$ metasurface, for 1550 nm x-polarized light incident at 45°, as a function of the volume fraction of metallic state $VO_2$ is shown in Figure 1b. As can be seen in Figure 1b, phase and amplitude cannot vary independently. This will result in angular side lobes angles in the far-field which has implications for phased array applications[31]. This antenna, optimized for a 180° phase shift, showcases a rather smooth phase shift with $VO_2$ composition and the minimum reflectance is not vanishingly small as observed when the total phase shift is maximized (when the resonance of the antenna coincides very precisely with the operating wavelength). The resonance phenomenon in each of the antenna is responsible for the phase shift can be seen in the field maps are shown in Figure 1c-d. A magnetic dipole resonance is evident when the $VO_2$ layer is in the cold, dielectric state. The field map is slightly asymmetric due to the 45° angle of incidence. This resonance disappears when the $VO_2$ transitions to its metallic state. The antenna state closest to resonance is obtained for 20% of metallic $VO_2$ and has the lowest reflectance. The high intrinsic losses in $VO_2$ lead to this problem, however the metasurface reflectance is higher when in a state further away from resonance. When the purely dielectric and purely metallic states have a reasonably high reflectance the antenna are more suitable binary control, as will be discussed further below.

Section II: Binary control

Binary control principle

Binary control has been proposed in the context of "programmable metasurfaces" or "coding metamaterials"[38]. It has been investigated experimentally for a few applications ranging from holography[39] to 5G phased arrays[40]. Binary or 1-bit control consists in switching each antenna into one of two states using an external stimulus. In this case the stimulus is Joule heating-to trigger $VO_2$'s IMT and induce a phase shift in the scattered electromagnetic field. We only consider the states where $VO_2$ is purely in the cold monoclinic state or in the hot rutile state which simplifies thermal control drastically. $d\Phi/dT$ is easily above 90°/K, see Figure 1b, so instead of precisely tuning the individual temperature of each element to a high precision, we can have a cold point well below the IMT transition temperature ($T_c$) and a hot point above $T_c$. The volatile transition of $VO_2$ enables full reconfigurability of the array and dynamic beam steering at high frequencies.

The angular dependence of the electric field for a reflecting array in the Fraunhofer conditions is given by equation 1. It is a direct summation of each antenna's complex contribution at every angle in space. The geometry and parameters are shown in Figure 2a. We designate as $\theta_i$ and $\theta_r$ the angle of incidence and desired anomalous reflection, $\Phi_m$ and $E_m(\theta)$ are the phase delay and the amplitude pattern of the E-field emitted by the $m^{th}$ antenna (out of a total of N) located in the 1D-array at position $x=x_m$ at an angle $\theta$, respectively.

$$E(\theta) = \sum_{m=1}^{N} E_m(\theta)\, e^{j(kx_m \sin(\theta_i) + kx_m \sin(\theta) + \Phi_m)} \quad 1$$

For beam steering applications, the electric field from all antennas have to constructively interfere at a given angle as to maximize energy in that direction. The generalized law of refraction[1] given in equation 2 in a homogenous medium gives the ideal phase shift profile for beam steering (derivation in Supplementary Information, section I).

$$\sin(\theta_i) + \sin(\theta_r) = \frac{\lambda}{2\pi} \frac{d\Phi}{dx} \quad 2$$

Integrating this equation gives a linear phase profile as in Figure 2b and d with a slope directly related to the anomalous reflection angle. It is directly applicable in ideal continuous phase shifting and requires 2π continuous phase shift capability. The binary control algorithm to convert this "ideal phase profile" is simple: we minimize the phase discrepancy between the binary phase shift profile and the ideal one. For Φ(x) ∈ [-π/2, π/2[, we use a phase shift of 0 ("OFF" state, the VO$_2$ element is in its cold dielectric state) and, for Φ(x) ∈ [π/2, 3π/2[, we have a π phase shift, the antenna is in "ON" state (the VO$_2$ element is in its hot metallic state). This selection algorithm is simple and maximizes by construction the power at angle $\theta_r$ but simultaneously maximizes the power sent at $-\theta_r$ for normal incidence, $\theta_i$=0. As can be seen in Figure 2b and c, binary control generates two symmetric beams which would be a limitation for LIDAR applications. However, it is technically possible to spatially filter out light in half of the hemisphere, but a better solution is to break this symmetry. These two beams correspond to sin($\theta_i$) - sin($\theta_r$) = ±α where α is a continuously tunable parameter in [-λ/2p ; λ/2p] corresponding to the right-hand term of equation 2. Using non-normal incidence, we can displace this $-\theta_r$ beam out of a region of interest defined by [-$\theta_{lim}$, $\theta_{lim}$]. The optimal angle of incidence maximizing the angular range of this region depends on the array period relative to the wavelength and is given in equation 3 (derivation in the SI, section II). Here we use p=λ/3 which corresponds to $\theta_{lim}$ = 48.6°. We round down to 45° to take the beam width into account and avoid a trailing edge of the $-\theta_r$ beam in the region of interest.

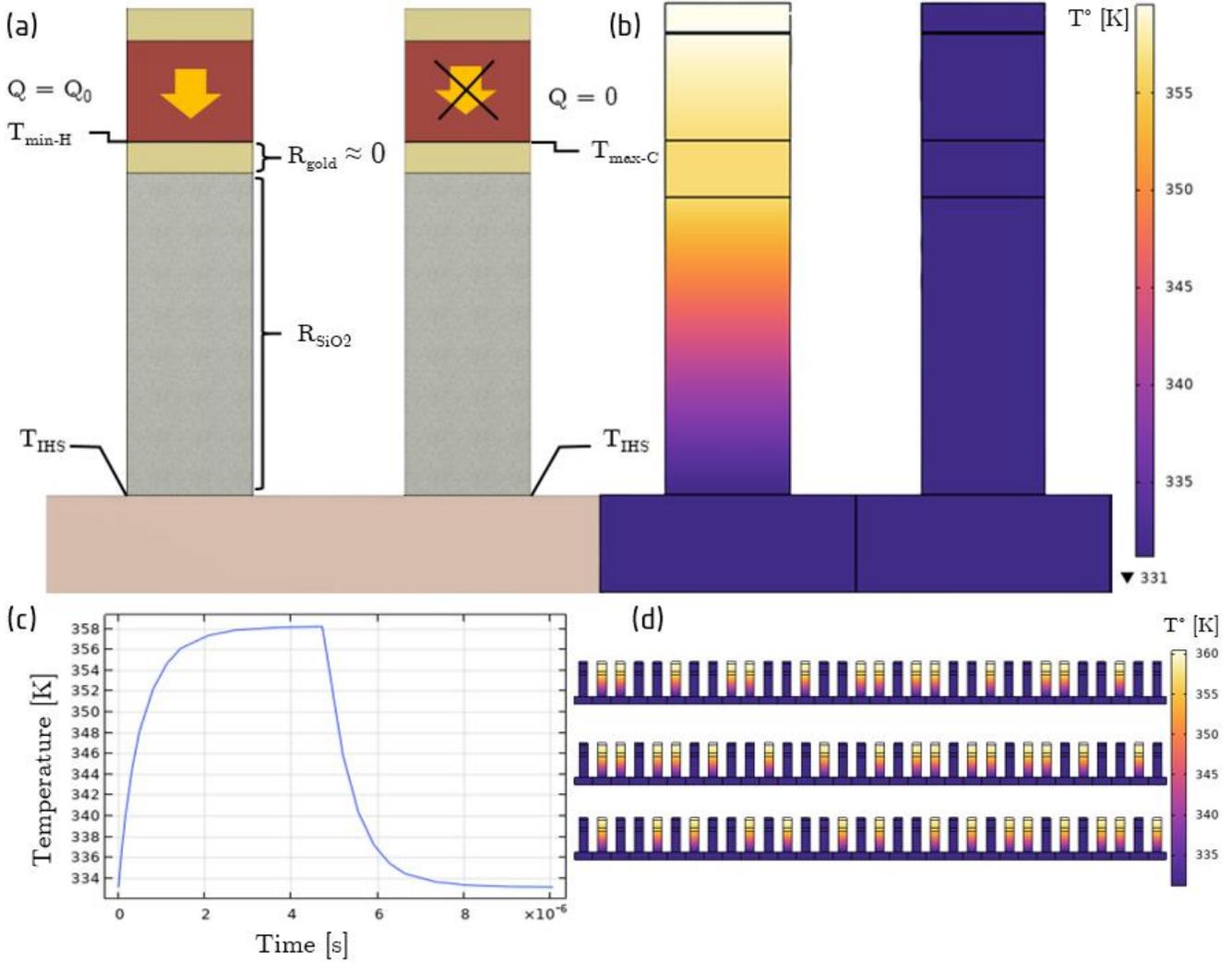

Figure 3: Thermal study  a - Heat transfer diagram of the antenna-scale problem  b - finite element study of the antenna-scale problem, steady state temperature distribution when only the left-antenna is turned ON  c - Transient behavior of the thermal contrast between two adjacent antennas in different states, 5µs heat pulse applied  d – Steady state temperature distribution of three configurations with N=32, $\theta_i$=45° (fixed) and varying values of $\theta_r$, respectively 10°, 20° and 30° (top to bottom), we demonstrate excellent thermal contrast between adjacent antennas for all patterns

$$\theta_{lim} = -\theta_i = \arcsin\left(\frac{\lambda}{4p}\right) \qquad 3$$

As the phase gradient shown in Figure 2d can be continuously tuned by the ON-OFF pattern. We can see in Figure 2e that it is possible to obtain any angle of anomalous refraction with binary control within [-$\theta_{lim}$, $\theta_{lim}$]. A given phase gradient usually results in an aperiodic antenna state arrangement, the number of possible arrangements specific to beam-steering scales with $N^2$ (where N is the number of antenna), faster than the number of resolvable points and not logarithmically as for periodic arrangements. Even for low values of N, the number of possible array-configurations greatly exceeds the number of resolvable points. Even if a discrete number of anomalous reflection angles is achieved, the beam steering is effectively continuous (see SI, section V). In the general case where the array is in a non-periodic arrangement, the beam intensity $I_{max} = I_{ideal}(2/\pi)^2$ (approximately 0.405 of the ideal case, derivation in the SI, section IV), this reduction in amplitude in the far-field is due to partially destructive interference induced by the discretization for binary control. Simulations show that for $\theta \approx \theta_r$, binary and continuous control have a similar beam shape to within a multiplicative constant. The FWHM remains the same as for continuous control and is given in equation 4[41]. We can see that the FWHM depends on the span (N*p) of the array relative to the wavelength, decreasing the array period will be counterproductive in that regard.

$$\Delta_{FWHM} = \frac{0.886\,\lambda}{N\,p\,\cos(\theta)} \quad\quad 4$$

Section II: Antenna design

Thermal design

Deposited Vanadium dioxide thin films are inhomogeneous structures that require a precise understanding of their microstructure to explain their optical behavior[32]. The film's microstructure grain boundaries and defect densities are the two main factors that dictate the thermal hysteresis and transition width[27]. In addition, adhesion to the substrate induces strain in the first few nanometers of the film which further complicates the thermal response. Hysteresis and transition width can be engineered up to a certain point. The MIM $VO_2$ array would benefit from a swift transition that requires the smallest thermal contrast between adjacent antennas, while ensuring one can be in the fully semiconducting $VO_2$ state while the other is in the fully metallic $VO_2$ state. We take a very conservative approach and design for a temperature contrast $\Delta T = T_{min-H} - T_{max-C} = 25°C$. This number could be potentially reduced depending on the $VO_2$ layer properties and other manufacturing or engineering factors.

As described earlier, the beam-steering metasurface under consideration is based on a MIM structure, with additional layers added for thermal purposes as shown in Figure 1d. Two thermal challenges have to be addressed at different scales. At the antenna-scale, thermal cross-talk must be prevented, this is not a trivial task as it requires maintaining a temperature difference of 25° over a few hundreds of nanometers. At the array scale, the heat generated by all the antennas must be dissipated. While the absolute power is relatively low (a few W), the heat flux is very high (~2000W/cm$^2$, see SI, section VI).

The antenna-scale heat problem is summarized in Figure 3a in a heat transfer diagram, and the detailed solution to this problem can be found in the SI (section IV). The minimum thermal insulator thickness necessary to ensure a thermal contrast $\Delta T$ can be calculated using equation 5

$$t_{insulator} = \frac{\Delta T\,k}{t_{Au}Q_{vol}} \quad\quad 5$$

where k is the thermal conductivity of the insulator and $Q_{vol}$ is the volumetric heat generation in W/m$^3$. This value is set to $9{,}67.10^{14}$ which corresponds to a current density of $2.10^{11}$ A/m$^2$, 10 times less than the experimental limit[42]. The $SiO_2$ thickness can be reduced by increasing the heat generation, the value of $t_{Au}$ = 60nm was found to be a good compromise between optical properties (which favors lower Au thickness as we will see later) and thermal performance (more heat generation with higher heater thickness). $SiO_2$ is a very convenient material choice as it is easily deposited and its low thermal conductivity of 1.4W/mK thins down the insulating layer. With these parameters, a $SiO_2$ insulating layer of 600nm is required to obtain $\Delta T = 25°$.

The steady-state temperature distribution in the array has been calculated using the commercially available finite element code COMSOL, as shown in Figure 3b. The IHS temperature is set to $T_{IHS}$ = 58°C = $T_c$-10°C, the convective heat transfer with ambient air is negligible at the antenna scale. The thermal contrast in the simulations agrees well with the calculations, the small discrepancy is due to the gold thickness whose thermal resistance and conduction in the substrate are neglected in our calculation. The transient behavior has similarly been modelled and is presented in Figure 3c, we find a very fast settling time, in the order of 2-3μs which could be expected given the high energy density in the device. Thermal simulations conducted at the array-scale are shown in Figure 3d with different binary control patterns, they correspond (top to bottom) to an anomalous reflection angle $\theta_r$ of 10°, 20° and 30°, respectively, for a fixed angle of incidence $\theta_i$=45°.

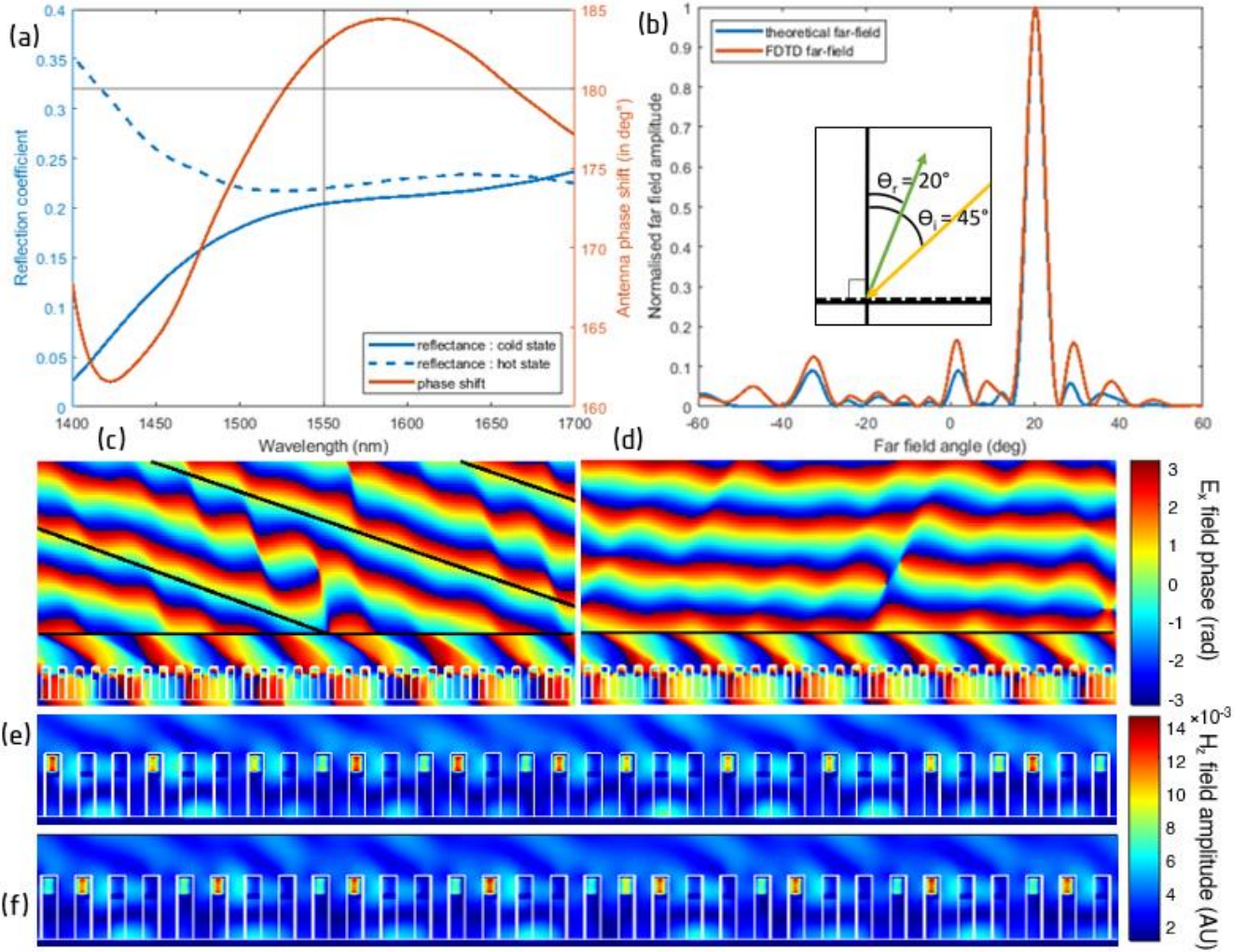

Figure 4: Array scale FDTD simulation results  a – broadband properties of the antenna, the reflectance in each state and the phase shift are shown between 1400 and 1700 nm, the slight difference at 1550nm with Figure 1-b arises from the interpolation of the material dispersion in Lumerical whereas a single wavelength was used for the simulation at 1550 nm in Figure 1   b - far field projection of the E-field obtained through FDTD simulations (red) compared to the direct theoretical E-field distribution from equation 1, incident angle $\theta_i=45°$, desired reflection angle $\theta_r=20°$. Inset, schematic showing the angles of incidence and reflection on the array   c/d - Ex field phase for $\theta_i=45°$, $\theta_r=20°$ and $0°$ respectively, we can clearly see beam steering even in the near field. The horizontal black line materializes the source position, we see the incident wave below the source and the reflected field above   d/f - Hz field amplitude for $\theta_i=45°$, $\theta_r=20°$ and $0°$ respectively, we can compare this figure to Figure 2d and easily see the antennas in the cold state exhibit a magnetic dipole resonance in the $VO_2$ element.

The array scale heat dissipation problem has already been studied intensively by microprocessor manufacturers[43], the use of Internal Heat Spreaders (IHS) is generally employed in the industry to cool down small components like this array. The idea is to spread the heat in a conductive plate to dissipate it over a larger surface area. An efficient IHS is necessary to cool down the array without resorting to more complex cooling methods, such as the use of cryogenics, liquids and enhanced forced convection, for example. This means a thermally conductive substrate is required, hence the use of a gold backplane in this structure. $SiO_2$ or other insulating materials are not suitable. The results of IHS model calculations (detailed in the SI, section IV) are shown Table 1 and compared to the finite element results, they validate the fact the array can be cooled down efficiently without resorting to advanced methods. $\Delta T_{avg}$ corresponds to the average temperature increase in the array and $\Delta T_{max}$ to the maximum temperature increase (usually at the center of the array where heat dissipation is the most difficult to achieve). For a fixed set of cooling conditions there will be an array size where the operating point will be above the $VO_2$ transition temperature, rendering it ineffective.

Table 1: Temperature increase: IHS model comparison to finite element simulation for an array-scale cooling problem

| Method | $\Delta T_{avg}$ | $\Delta T_{max}$ |
|---|---|---|
| IHS model | 31.87 | 32.06 |
| Finite element | 31.54 | 31.77 |

Inverse design

Now that we have set several design parameters to obtain a functional binary controlled phased array, we can optimize the structure to maximize its performance while respecting the aforementioned engineering constraints. To conduct this multi-parameter optimization, we employ inverse design. Machine learning has opened new possibilities in many fields, its applications in photonics are just starting to be explored[25]. Inverse design lets an algorithm adjust some degrees of freedom (DOF) to optimize a user-designed figure of merit (FOM). This allows for multiple DOF simultaneous optimization. There are many algorithms that can be used to optimize a structure. Given the specifics of this case the hybrid PSO – interior-point algorithm is the most relevant (more details can be found in the SI, section VIII). The gold heater thickness is fixed as the inverse design algorithm finds better performance with lower thickness values that are incompatible with our thermal constraints. A value $t_{Au\ thermal}$ = 60 nm is a good compromise between energy consumption, thermal contrast and reflectance efficiency. The only values that are optimized here are therefore the antenna width w, the VO$_2$ thickness $t_{VO2}$ and the gold reflector thickness $t_{Au\ reflector}$.

Table 2: Antenna design parameters values

| Variable | Value (nm) |
|---|---|
| p | 516.7 (fixed to λ/3) |
| w | 245.2 |
| $t_{SiO2}$ | 600 (fixed) |
| $t_{VO2}$ | 216.3 |
| $t_{Au\ thermal}$ | 60 (fixed) |
| $t_{Au\ reflector}$ | 99.4 |
| $t_{Ti}$ | 2 (fixed) |

Binary control is ideally implemented with a maximized phase contrast of π while maintaining the amplitude ratio of 1 between the two antenna states. The FOM given in equation 6, is calculated from two simulations carried out on the same geometry, with the VO$_2$ in dielectric and rutile state. The FOM is the product of two terms that have to be maximized simultaneously.

$$FOM = gauss(\pi, \pi/9).\min(R_{ON}, R_{OFF}) \quad 6$$

The first term is a Gaussian centered at 180° with a standard deviation of 20° (arbitrary value) to maximize the phase contrast between both states with vanishing values when the phase shift is far from π. The second term is the minimum reflectance of both states, this incentivizes the algorithm to increase only the state with minimum reflectance to reduce the amplitude discrepancy. Eventually this FOM component also increases the overall reflectance but unlike any other formula (geometric mean or arithmetic mean for example), it does not push the algorithm to increase the reflectance of one state at the expense of the other. Note that it is better for our array performance to have a 0-reflectance discrepancy rather than a higher reflectance in a single state to avoid side lobes. The hybrid inverse design algorithm results can be found in Table 2, these are optimal values as defined by our FOM and within the engineering bounds we have specified. As it is a versatile algorithm, we also show results in the SI for alternative geometries that could correspond to other engineering choices and their associated performances. The optimization algorithm can be applied to finding structures suitable for different criteria such as operating wavelength, material properties or even applications. Examples of geometries optimized for different parameters are shown in the SI (section VIII).

Section III: Antenna performance

The geometry prescribed in Table 2Error! Reference source not found. has been assessed thoroughly using the commercially available finite difference time domain software Lumerical. As reported by other studies, this resonance causes a phase shift in the reflected beam but also decreases the reflectance to 21% due to the strong light-matter interaction in dissipative materials, as seen in Figure 1b. The broadband performance of the array is assessed and presented in Figure *4*a. The metasurface, despite the resonant behavior of each antenna maintains its performance over a wide band, especially above its operating wavelength of 1550 nm for up to 100 nm.

Now that the antenna geometry is optimized, the metasurface is programmed using binary control with some antennas turned ON and some others turned OFF. The results for the FDTD simulation of a full array with N=32 elements, an angle of incidence θ$_i$=45° and the desired reflection angle θ$_r$=20° is presented in Figure 4b. The far-field pattern obtained from the simulations is also compared to first-principles calculations using equation 1. One can see in Figure 4c-d the beam steering patterns and more specifically in Figure 4e-f the magnetic dipole resonance in each antenna in the cold state.

Excellent agreement of the theoretical model with numerical simulations is obtained demonstrating that electromagnetic crosstalk remains limited, and the antennas performance is unchanged when placed in an array with adjacent antennas in a different state. Given the broadband response of the metasurface a scheme employing wavelength multiplexing to scan several angles at once could be considered. The antenna behaves close to a perfect binary antenna with π phase shift and near-unity amplitude ratio of the reflectance in the cold and hot states over a broad wavelength band. If the incident beam contains multiple wavelengths the anomalous reflection will separate them, which could be used to increase the scanning speed, for example in LiDAR applications. This could enable another degree of freedom to steer the beam around a second axis by tuning the illumination wavelength[44].

Binary control can be applied without any loss of generality to any phased array regardless of the mechanism used to tune the output phase. For example, this approach can be applied to phased arrays using other PCMs or external phase shifters at any wavelength and scale. All the analytical derivations have been presented in 1D but can easily be extended to 2D metasurfaces (see SI section III). Furthermore, one can note that we have only considered a structurally periodic (with uniform antenna spacing) phased array with uniform illumination (all the antennas are subjected to the same incident field amplitude), this kind of device is not known for its optimal performance. Structurally aperiodic arrays tend to perform better in practice[45] and amplitude tapering is also very useful to reduce side-lobes intensity. We did not extend the analysis to these special phased arrays in order to keep the analysis as reproducible and general as possible but performance in applications can be further improved using these concepts. While structural aperiodicity may prove be more taxing to achieve, as changing the gap between adjacent antennas may influence their response, amplitude tapering is almost guaranteed as the illumination from a laser source has a non-constant beam amplitude profile by nature. A higher thermal contrast could also be implemented in a 2D phased array to enable multiplexing, carefully taking advantage of the thermal inertia of the device. Array-level inverse design[46] could also be implemented to improve on the binary control algorithm with a possibility to enhance user-defined properties of the far-field. Finally, the simple inverse design algorithm exploited to tune the antenna parameters could be used with another set of engineering constraints or FOM in order to get a different functionality.

Conclusion

We have successfully applied binary control to phase change nanoantenna arrays. It is seen that this approach provides a solution to many of the issues that have been encountered in the practical implementation of tunable resonant phased array metasurfaces in the NIR and optical domains. The performance decrease compared to ideal continuous control is compensated by an easy implementation based on a control of the binary state of each antenna. This approach can be applied to tunable metasurfaces for a wide range of applications. A thermally realistic $VO_2$ based MIM antenna has been investigated. Using binary control, combined, with inverse design at the antenna level within feasible engineering limits, broadband continuous beam-steering over a 90º angular range between 1500 and 1700 nm has been demonstrated. It is expected that inverse design and machine learning at the antenna and array levels will reveal designs capable of high performance coupled with less demanding implementation requirements for a wide range of metasurface wavefront engineering challenges and applications.

ASSOCIATED CONTENT -

Derivation and proof of equations, detailed heat transfer calculations and more simulation results are available in the Supplementary Information. This material is available free of charge via the Internet at http://pubs.acs.org

## AUTHOR INFORMATION

### Corresponding Author

A. Louise Bradley - School of Physics and AMBER, Trinity College Dublin, Dublin 2, Ireland;  orcid.org/0000-0002-9399-8628
*Email: bradlel@tcd.ie

### Authors

Matthieu Proffit - School of Physics and AMBER, Trinity College Dublin, Dublin 2, Ireland
Sara Peliviani - School of Physics and AMBER, Trinity College Dublin, Dublin 2, Ireland
Pascal Landais - School of Electronic Engineering, Dublin City University, Glasnevin, Dublin 9, Ireland

### Author Contributions

The manuscript was written through contributions of all authors. All authors have given approval to the final version of the manuscript.

### Funding Sources

This work was funded by Science Foundation Ireland (SFI) under grants 16/IA/4550 and 12/RC/2278_2.

## ACKNOWLEDGMENT

The authors thank C. Smith for the ellipsometry measurements